\documentclass[amssymb,amsmath,nobibnotes,aps,prl,preprint,groupedaddress,showpacs]{revtex4}

\usepackage{graphicx}

\begin{document}


\title{Interference of a variable number of coherent atomic sources}

\author{Giovanni Cennini}
\author{Carsten Geckeler}
\author{Gunnar Ritt}
\author{Martin Weitz}
\affiliation{Physikalisches Institut der Universit\"{a}t T\"{u}bingen, Auf der
Morgenstelle 14, 72076 T\"{u}bingen, Germany}

\date{\today}

\begin{abstract}
We have studied the interference of a variable number of
independently created $m_F=0$ microcondensates in a CO$_{2}$-laser
optical lattice. The observed average interference contrast
decreases with condensate number $N$. Our experimental results agree well
with the predictions of a random walk model. While the exact result
can be given in terms of Kluyver's formula, for a large number of
sources a $1/\sqrt{N}$ scaling of the average fringe contrast
is obtained. This scaling law is found to be of more general 
applicability when quantifying
the decay of coherence of an ensemble with $N$ independently
phased sources.
\end{abstract}


\maketitle 
Interference experiments are valuable tools for the study 
of coherence of atomic sources~\cite{Andrews, Bloch}. 
While for an
incoherent ensemble the number of independently phased objects
equals the atom number, in a Bose-Einstein condensate atoms 
macroscopically populate a one-particle quantum state and 
the whole condensate is a single coherent atom source. Of large interest is 
the intermediate region of partial coherence. 
We shall here be 
interested in the degree of coherence of an ensemble of $N$ 
independent quantum objects. 
For an array of periodically
spaced atom sources, each of which be coherent, Ashhab predicted a $1/\sqrt{N}$ scaling
of the average fringe visibility with the number of sources $N$ 
in a far field interference experiment \cite{Ashhab}.
For other theoretical works, see~\cite{Lattice-Theory}.
Experimentally, Hadzibabic et al.~\cite{Hadzibabic}
had observed the interference from 30 independent condensates 
and obtained the surprisingly high average
fringe contrast of $34\,\%$, which was indeed comparable to the
results observed in the early two-condensates interference
experiment of~\cite{Andrews}. The experimental result of
Hadzibabic et al. was in agreement with a Monte Carlo simulation. 

We here experimentally investigate the interference of a
variable number of randomly phased atomic 
sources. To verify the predicted $1/\sqrt{N}$ scaling
of the fringe visibility as a function of the number $N$ of sources, 
an array of magnetic field-insensitive ($m_{F}=0$) Bose-Einstein 
condensates is created by all optical cooling and trapping techniques.
In the sites of a one-dimensional mesoscopic 
optical lattice potential, rubidium atoms are independently cooled
evaporatively to quantum degeneracy. After releasing the 
microcondensates, an interference pattern is created. The
observed average fringe visibility decreases with the number of
coherent sources over which the atoms are distributed. Our results well follow the predictions of an exact theoretical
model based on a random walk in the complex plane. In the limit of
large number of condensates $N$, the expected $1/\sqrt{N}$ scaling of the 
average fringe visibility characteristic for the central limit result 
of the random walk is experimentally reproduced. 

The observed variation of fringe visibility with the number of sources can 
intuitively be understood as follows. When two independent condensates with
equal number of atoms are overlapped, upon 
measurement the condensates 
are projected onto a state with well-defined relative phase.
This phase is random in each realization of the
experiment, but in all cases a density modulation with $100\,\%$
contrast occurs~\cite{Javanainen,Naraschewski,Castin}.
For three interfering
condensates now two relative phases come into play, and in general
the phase difference between any two of the condensates differs
from that between either one of them and the third condensate. One
still expects a periodic density modulation, but with reduced
contrast. For a larger number of independent sources, the
degree of randomness increases. The fringe contrast 
for an array of $N$ coherent sources can 
be calculated with a random-walk model in which the number of steps
equals the number independent phases, i.e.~$N-1$ \cite{Ashhab}.

We wish to point out that the predicted $1/\sqrt{N}$ scaling of the fringe
visibility with the number of sources $N$ in the lattice is actually a far
more general result when considering the degree of coherence of 
atomic ensembles.  As a textbook-like example, let us consider
a sample with $N$ independent monochomatic sources (single atoms in 
a classical model or independent condensates) with total wavefunction 
$\Phi (\textbf{x}) \propto
\sum^{N}_{n=1} e^{i(\textbf{k}\cdot\textbf{x} +\theta_{n})}$, where the phases
$\theta_{n}$ ($n=1,\ldots,N$) are randomly distributed. To test for the
coherence of this ensemble, we shall beat the sample with a coherent local
oscillator wave (e.g.  an atom laser) with wavefunction
$\Phi_{\mathrm{lo}}(\textbf{x}) \propto e^{i \textbf{k}_{\mathrm{lo}}\cdot
  \textbf{x}}$.  This yields a spatial interference signal 
$|\Phi+\Phi_{\mathrm{lo}}|^2$ with intensity
modulation: $\Phi_{\mathrm{lo}}^{*} \Phi + \mathrm{c.c.}$ =
$2\,\mathrm{Re}\,(e^{i (\textbf{k}-\textbf{k}_{\mathrm{lo}})\cdot\textbf{x}}
\sum^{N}_{n=1}e^{i\theta_{n}})$.  After evaluating the average value of this
random walk problem with $N$ steps, one finds that the modulation of the
fringe signal scales as $\sqrt{N}$, which compares to the value of $N$ that is
obtained with a coherent atomic ensemble.  The visibility of the interference
signal, a quantity commonly used when quantify coherence in the optics
literature \cite{optics}, decays with a $1/\sqrt{N}$ scaling law, i.e.
relatively slowly with the number of sources.

In our experimental approach, atoms are evaporated to quantum degeneracy in
the sites of a one-dimensional periodic potential created by an
optical standing wave generated by CO$_2$-laser whose wavelength
is near $10.6\,\mbox{$\mu$m}$. This laser frequency is roughly a
factor 14 below the lowest electronic resonance frequency of the
rubidium atoms used in our experiment. Due to the static atomic
polarizability, atoms are pulled into the antinodes of the
standing wave. Adjacent sites are spaced by
$d=\lambda_{\rm{CO_{2}}}/2\simeq5.3\,\mbox{$\mu$m}$, which is
sufficiently far apart that tunneling between sites is completely
negligible. A variable number of sites is initially occupied with
thermal atoms, which are subsequently cooled to quantum degeneracy
by direct evaporation in the optical trap. 
Our experiment benefits
of recent developments in the ``all-optical'' generation of
Bose-Einstein condensates in optical dipole
potentials~\cite{Barret,Weber,Takasu,Cennini}, which we here
extend to a lattice geometry. Besides offering the
flexibility to directly evaporate into the periodic atom potential,
in such traps magnetic-field insensitive
condensates can be produced~\cite{Takasu,Cennini}, which is a
clear benefit in studies of coherence properties.

Let us theoretically discuss the expected interference signal
after releasing the array of $N$ (with $N$ variable) independent
microcondensates in the far field. This treatment is
inspired by earlier calculations considering the interference of independent
optical~\cite{Merzbacher} and atomic~\cite{Ashhab,Hadzibabic} sources. 
In our model, we consider the $N$
independent atomic Bose-Einstein condensates with same trap
parameters in all sites before the free expansion. The total
wavefunction at this time can be written as
\begin{equation}\label{eq1}
\Phi(x)=\sum^{N}_{n=1}{a_{n}e^{i\theta_{n}}e^{-(x-x_n)^2/(2l^2)}}
\end{equation}
where $a_{n} e^{i\theta_{n}}$ denotes the amplitude of the $n$-th
condensate wavefunction, which is assumed to be a gaussian
function with ground state size $l=\sqrt{\hbar/(m\omega)}$. The
microcondensates are centered at the CO$_2$-laser lattice
antinodes, corresponding to the positions $x_{n}=n\cdot d$, where
$n=1,\ldots,N$. Because the microcondensates are independent, the
phase factors $\theta_{n}$ are randomly
distributed~\cite{Javanainen,Naraschewski,Castin}. We can immediately derive the wave
function in a momentum space picture $\tilde{\Phi}(p)$ by
performing a Fourier transform. When expanding the atomic clouds
and recording the interference pattern with a time of flight
measurement in the far field, we actually perform a measurement of
$|\tilde{\Phi}(p)|^2$, which can be written as
\begin{equation}\label{eq2}
\bigl|\tilde{\Phi} (p)\bigr|^2=\frac{l^2}{\hbar}\biggl[\,\sum^{N-1}_{n=1}
{A_{n}\cos(ndp/\hbar+\varphi_{n})}\,\biggr] \,e^{-p^2l^2/\hbar^2}.
\end{equation}
In this equation, $A_0 = N a^2$, $\varphi_0=0$. For $n$ other than
zero the (real) amplitudes $A_n$ and the corresponding phase
angles $\varphi_n$ are given by~\cite{Ashhab} $A_n
e^{i\varphi_n}=2 a^2
\sum^{N-n}_{q=1}e^{i(\theta_{q}-\theta_{q+n})}$,
if we for the sake of simplicity assume equal atom population for all sites (i.e.~$a_n=a$ for
all $n$).

In our experiment, we only spatially resolve the fringe pattern
with largest spatial modulation, i.e. the interference between
nearest neighbours. The expected visibility of this modulation is
given by $V\equiv
(I_{\rm{max}}-I_{\rm{min}})/(I_{\rm{max}}+I_{\rm{min}}) =
A_1/A_0$, which we can write more simply as $V = 2 S/N$, where $S=
|z|=\bigl|\sum^{N-1}_{q=1}z_{q}\bigr|=\bigl|\sum^{N-1}_{q=1}e^{i\Delta\theta_{q}}\bigr|$
and $\varphi_{1}\equiv\varphi$ equal the phase angle of this sum.
The phase angles $\Delta\theta_{q}$ are randomly distributed in
the range $0\leq\Delta\theta_q<2\pi$. For each realization of the
experiment, the fringe contrast is determined by the modulus of a
sum of unit vectors $z_{q}$ which have random directions in the
complex plane. Further, the phase angle of the total vector gives
the position of the fringe pattern. In our model,
the problem of calculating the fringe contrast and
position of an array of coherent, independent sources thus can be 
transferred to the solving of a random walk problem in the complex
plane. This is a typical two-dimensional Pearson's random
walk~\cite{Pearson,Weiss}. Fig.~1a gives a graphical
representation of the situation. In general, the expected
probability density to arrive at a distance $S$ from the origin is
given by Kluyver's formula~\cite{Kluyver}:
\begin{equation}\label{eq4}
p(S,N-1)=\frac{1}{N-1}\int^{\infty}_{0}{\bigl[J_0(\rho)\bigr]^{N-1}J_0(S\rho)\rho\,
d\rho}
\end{equation}
where $J_0$ is the 0-th Bessel function. For large $N$, the Bessel
functions can be asymptotically expanded, and a Gaussian
probability distribution is obtained, corresponding to the
celebrated central limit result~\cite{Weiss}. In this
approximation, the expected average visibility approaches the
previously derived value of $\sqrt{\pi/(N-1)}\simeq\sqrt{\pi/N}$,
when $N\gg1$~\cite{Ashhab}. Note that $N-1$ equals the
number of relative phases between adjacent sites and therefore the number of steps of the
random walk. In the general case, and especially for a small
number of condensates, one has to fully solve Eq.~\ref{eq4} which
is difficult due to the oscillatory character of the Bessel functions. We
have numerically evaluated the probability density by expanding
Kluyver's formula in a Bessel-Fourier series~\cite{Merzbacher}. As
an independent verification of our calculation, we performed a
Monte Carlo analysis of the average visibility assigning random
values to the arguments $\Delta\theta_{q}$. Fig.~1b shows the theoretical
expected fringe average visibility as a function of condensate
number derived using the exact calculation (dots), the central
limit result (dashed line) and the simple $\sqrt{\pi/N}$ scaling
function (dashed-dotted line).

An array of independent Bose-Einstein condensates with variable
number of sites is produced using a modification of our previous
setup~\cite{Cennini,C_ApplPhys}. In an ultrahigh vacuum chamber,
mid-infrared radiation near $10.6\,\mbox{$\mu$m}$ for atom
trapping is focused to a beam waist ($1/e^2$ radius) in the range
$25$--$40\,\mbox{$\mu$m}$. With a second lens and a
retroreflection mirror, a standing wave is generated. The size of
the beam waist in the trapping region could be varied with a
telescope. This gave us control over the number of populated
trapping sites. For a loading of atoms, we switched to a running
wave geometry by slightly misadjusting the CO$_{2}$-laser beam
backreflection mirror away from a perfect retroreflection. This
mirror was mounted on a piezo, which allowed for an electronic
control of the degree of alignment during the course of the
experiment. Cold rubidium ($^{87}$Rb) atoms from a magneto-optical
trap (MOT) were loaded into the running wave geometry. After this
transfer, 10$^{6}$ atoms, populating the lower hyperfine ground
state ($F = 1$, $m_F=0,\pm1$) are left in the optical trap at a
temperature near $100\,\mbox{$\mu$K}$. To cool the trapped atomic
cloud, the CO$_2$-laser beam power is acousto-optically ramped
down to induce forced evaporative cooling. The highest energetic
atoms leave the purely optical trap, and the remaining ones
thermalize to lower temperatures. The total evaporation stage
lasts about 10 seconds, during which the mid-infrared beam power
is smoothly reduced from $30\,\rm{W}$ to a typical final value of
$40$--$50\,\rm{mW}$. Throughout the evaporation stage, the MOT
quadrupole field with $10\,\rm{G}/\rm{cm}$ field gradient is left
on. This gradient is sufficiently strong to remove atoms in
field-sensitive spin projections in this running wave geometry, as
the trapping force along the beam axis is here relatively
weak~\cite{Magnetic}. While the initial phase of this evaporation
is still performed in the running wave geometry (this limits the
number of sites over which atoms are distributed), during the
course of the evaporative cooling we slowly switch to a standing
wave (lattice) geometry by servoing the piezo-mounted mirror
correspondingly. The 1D lattice geometry is fully aligned at a
time 1--2 seconds before the onset of quantum degeneracy
(corresponding to atomic temperatures of a factor~2 above the
transition temperature of about $T_{\rm{c}}\simeq200\,\rm{nK}$).
This ensures that the microcondensates are formed independently.

At the end of the evaporation state, an array of $m_F=0$
microcondensates is created. The total number of atoms in the
optical lattice is about 7000, and the number of condensed lattice
sites can be varied from typically 5 to 35 by choosing different
beam waists of the lattice beams and -- in a smaller range -- also
allowing for tiny residual misalignments of the lattice beams at
given beam diameter. For all our measurements, the estimated
tunneling time between sites is above $10^{18}\,\rm{s}$. This
assures that the microcondensates are truly independent. The
residual sensitivity of the $m_F=0$ condensates to stray magnetic
fields due to the second order Doppler shift is near
$14\,\rm{fK}/\rm{mG}^{2}$. For a typical extension of the lattice
of $100\,\mbox{$\mu$m}$, the variation of the condensate phases
due to magnetic field inhomogenities is $0.15\,\rm{Hz}$ at an
estimated field gradient of $50\,\rm{mG}/\rm{cm}$. On the other
hand, we expect that differential mean field shifts due to
variation in the atom number do cause a non-negligible
differential variation of the condensate phase with time.

After creation of a variable number of microcondensates in the
lattice, the CO$_{2}$-laser radiation is extinguished, and the
atomic clouds are allowed to freely expand and fall in the earth's
gravitational field. The condensed atomic clouds originating from
adjacent sites overlap as soon as the free expansion time $t$
exceeds $0.5$--$1.5\,\rm{ms}$, depending on the interaction energy
for the particular atom population per site, which in turn depends
on $N$. We then expect to observe an interference pattern with
spatial period of $\lambda_{\rm{th}}$ = $ht/(md)$, despite each
individual atom source having random phase. Figure~2a shows a
typical observed interference pattern for $N=20$ interfering
coherent atomic clouds. The free expansion time here was
$t=15\,\rm{ms}$, after which the shown absorption image was
recorded. A horizontal profile of this image is shown in Fig.~2b,
where the dashed line with dots shows the experimental data. For an
analysis of our data, the fringe patterns (i.e. the
absorption image profiles) were fitted with the function
\begin{eqnarray}\label{eq5}
I(x) &=&
I_{0}\bigl[e^{-(x-x_{0})^{2}/{\sigma}^2}+V_{\exp}e^{-(x-x_{0}^{\prime})^{2}/\sigma^{\prime
    2}}\\\nonumber
& & {}\cdot \cos(2\pi x/\lambda_{\rm{exp}}+\varphi_{\rm{exp}})\bigr],
\end{eqnarray}
where $\lambda_{\rm{exp}}$ and $\varphi_{\rm{exp}}$ denote the
fitted fringe spacing and the phase angle of the 
interference pattern. Gaussian envelopes with independent widths and
positions are here included for the background and fringe signals 
respectively.
The fringe visibility $V_{\rm{exp}}$ of the pattern of Fig.~2a, as
derived from the fit, was $31.5\,\%$. The experimental fringe
spacing here is
$\lambda_{\rm{exp}}\simeq13.0\pm0.3\,\mbox{$\mu$m}$, which agrees
well with the expected spacing
$\lambda_{\rm{th}}\simeq12.97\mbox{$\mu$m}$. We have studied the
variation of fringe contrast and phase angle for different
realizations of the experiment. Fig.~2c shows these parameters in
a polar plot, where each dot corresponds to the result of an
individual realization. Phase angles and fringe contrasts here
vary randomly from measurement to measurement. Moreover, no
preference for a certain value of phase angle is visible. Note the
close analogy of this polar image to Fig.~1a. The vector
$z=Se^{i\varphi}$ as a complex signal modulation (where
$S=\frac{N}{2}V$) represents the endpoint of a random walk in the
complex plane, and can be directly experimentally determined from
fringe contrast and phase of a far field interference pattern.
Series of fringe patterns were recorded for different numbers $N$
of populated lattice sites.

Fig.~3 shows the variation of the average fringe visibility for a
series of measurement as a function of $N$. The average number of
populated sites here was determined by recording the length of the
lattice, as measured with a time of flight image at zero expansion
time. Our experimental data clearly shows a decrease of the fringe
contrast with condensate number, which agrees with the exact
random walk solution valid for arbitrary $N$ (see Eq.~\ref{eq4})
and within our experimental uncertainties also with the central
limit result. For large number of independent sources, the fringe
contrast averages out. However, this averaging effect occurs
relatively slowly, i.e. it takes a large number of sources to let
the fringe pattern resemble that of an incoherent atomic sample.
The data shown in Fig.~3 has been fitted with the (interpolated)
result of the exact calculation, as shown by the solid line. To
account for our experimental imaging resolution, we have
multiplied the theoretical result with a constant value, which was
the only free parameter in the fit. The resolution of
$6.3\,\mbox{$\mu$m}$ obtained in this way agrees well with an
independent, direct measure of the optical resolution of
$6.0\,\mbox{$\mu$m}$. The dashed line gives the corresponding
visibilities using the simple $\sqrt{\pi/(N-1)}$ result of the
central limit approximation, which was multiplied with the same
constant factor.

To conclude, we have studied the interference of an array of
independently generated $m_F=0$ microcondensates in an optical
lattice for a variable number $N$ of populated sites. 
The observed average interference visibility decreases with
source number due the increased number of
(upon measurement) randomly distributed relative phases,
and well follows the predicted $1/\sqrt{N}$ scaling.

For the future, we anticipate that the observed relatively 
slow decay of coherence with source number will allow
one to observe related interference effects with small
samples of thermal atoms.
A further direction for future research would be to
experimentally verify the intrinsic (i.e. quantum-mechanical) 
randomness of the relative phases of the atom sources. 
Due to the use of $m_F=0$ condensates, differential phase 
shifts due to stray magnetic fields are estimated to contribute 
clearly below a radian.
Our current experiment, differential mean field phase shifts 
due to the in general slightly different atom numbers per site 
during the approximately
1-2 seconds time period between the onset of quantum degeneracy
and the release of the microcondensates are however estimated to still
accumulate statistical variations of the phase $\varphi$ of
several tens of radians from shot to shot. If e.g. measurements with
constant atom number per site could be postselected, which clearly 
requires small absolute atom numbers, the intrinsic
randomness of the condensates relative phase might be
experimentally tested. Finally, we anticipate that our
experimental scheme holds promise for novel types of lattice atom
lasers.

We acknowledge financial support from the Deutsche
Forschungsgemeinschaft, the Landesstiftung Baden W\"urttemberg,
and the European Community.

\small{


\newpage
\begin{figure}
\caption{(a) Random walk of coefficients
$z_{q}=e^{i\Delta\theta_{q}}$, with the phase angles
$0\leq\Delta\theta_{q}<2\pi$ arbitrary in the complex plane. The
vector $z$ is the sum of $N-1$ independent coefficients $z_{q}$
($1\leq q\leq N-1$) and determines both the fringe visibility
$V=2S/N$ with $S=|z|$ and its phase angle $\varphi$. (b) Predicted
average fringe visibility for the far field interference pattern
of $N$ coherent atomic sources (dots). Result of the central limit
approximation: $V=\sqrt{\pi/(N-1)}$ (dashed line) and the simple
$V=\sqrt{\pi/N}$ scaling function (dashed-dotted line).}
\end{figure}
\begin{figure}
\caption{(a) False-color absorption image of an interference
pattern of 20 independently generated $m_F=0$ microcondensates
released from a CO$_{2}$-laser optical lattice. The image was
recorded after a free expansion time of 15 ms. The field of view
is $350\,\mbox{$\mu$m}\times140\,\mbox{$\mu$m}$. (b) Horizontal
profile of the image averaged over a vertical region of
$35\,\mbox{$\mu$m}$: Experimental data (solid line with dots) and
fitted fringe pattern (dashed line). (c) The fringe visibilities 
and phase angles of fringe patterns
arising from the interference of 20 independent Bose-Einstein
condensates, as derived from fits as in Fig.~2b, have been drawn
(dots) in a polar diagram for 130 different realizations of the
experiment.}
\end{figure}
\begin{figure}
\caption{Average fringe visibilities of far field interference
patterns as a function of number of interfering coherent atomic
sources. The experimental data sets were recorded with beam
waists: $24.3\,\mbox{$\mu$m}$ (squares), $30.0\,\mbox{$\mu$m}$
(dots), and $40.2\,\mbox{$\mu$m}$ (circles). The data has been
fitted with the (interpolated) theoretical fringe contrast
multiplied by a constant factor, which is left as a free parameter
to account for the finite imaging resolution
(solid line). The dashed line gives the corresponding result in
the central limit approximation when assuming
the same imaging resolution.}
\end{figure}


\begin{thebibliography}{10}


\bibitem{Andrews}
M. R. Andrews {\it et al.}, Science \textbf{275}, 637
(1997).

\bibitem{Bloch}
See e.g: I. Bloch, Phys.\ World \textbf{17}, 25 (2004).

\bibitem{Ashhab}
S. Ashhab, cond-mat/0407414.

\bibitem{Lattice-Theory}
A. R. Kolovsky, Europhys.\ Lett.\ \textbf{68}, 330 (2004); R. Bach
and K. Rzazewski, cond-mat/0407022.

\bibitem{Hadzibabic}
Z. Hadzibabic {\it et al.}, Phys.\ Rev.\ Lett.\ \textbf{93},
180403 (2004).

\bibitem{optics}
See e.g: E.G. Steward, {\it Fourier Optics}, (Ellis Horwood, Chichester, 
1989).

\bibitem{Javanainen}
J. Javanainen and S. M. Yoo, Phys.\ Rev.\ Lett.\ \textbf{76}, 161
(1996).

\bibitem{Naraschewski}
M. Naraschewski {\it et al.}, Phys.\ Rev.\ A \textbf{54}, 2185
(1996).

\bibitem{Castin}
Y. Castin and J. Dalibard, Phys.\ Rev.\ A, \textbf{55}, 4330 (1997).

\bibitem{Barret}
M. Barrett, J. Sauer and M. S. Chapman, Phys.\ Rev.\
Lett.\ \textbf{87}, 010404 (2001).

\bibitem{Weber}
T. Weber {\it et al.}, Science \textbf{299}, 232
(2003).

\bibitem{Takasu}
Y. Takasu {\it et al.}, Phys.\ Rev.\ Lett.\ \textbf{91}, 040404
(2003).

\bibitem{Cennini}
G. Cennini, G. Ritt, C. Geckeler, and M. Weitz, Phys.\ Rev.\
Lett.\ \textbf{91}, 240408 (2003).

\bibitem{Merzbacher}
See e.g.: E. Merzbacher, J. M. Feagi, and T.-H. Wu, Am.\ J. Phys.\
\textbf{45}, 964 (1977).

\bibitem{Pearson}
K. Pearson, Nature, Vol.\ LXXII, 294 (1905).

\bibitem{Weiss}
See e.g: G. H. Weiss, {\it Aspects and Application Of The Random Walk}
(North-Holland, 1994).

\bibitem{Kluyver}
J. C. Kluyver, Konink. Acad.\ Wetenschap te Amst.
\textbf{14}, 325 (1906).

\bibitem{C_ApplPhys}
G. Cennini, G. Ritt, C. Geckeler, and M. Weitz, Appl.\ Phys.\ B \textbf{77}, 773 (2003).

\bibitem{Magnetic}
The center of the MOT quadrupole field is in general not perfectly
aligned with the optical dipole trap.

\end{thebibliography}
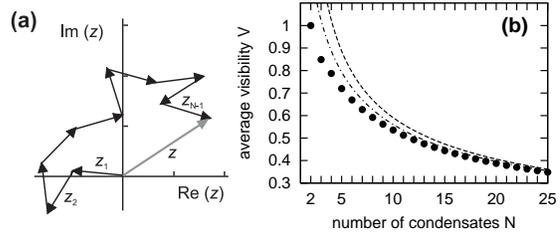
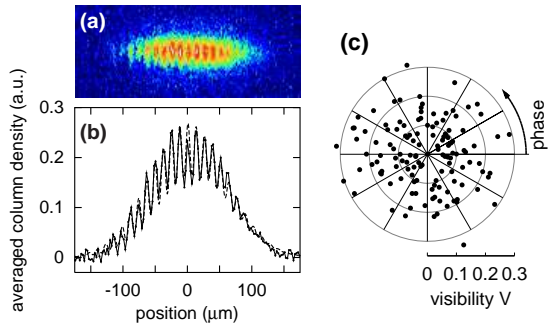
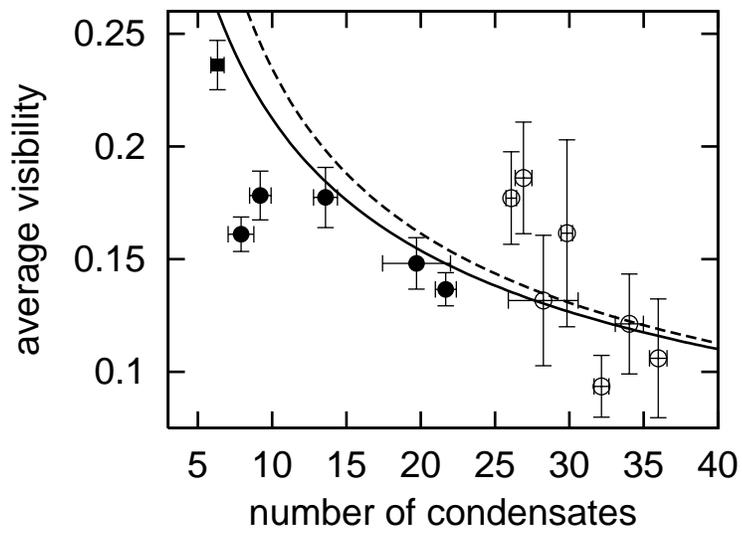
\end{document}